\documentstyle[proceedings,epsf]{crckapb}

\begin{opening}

\title{The Variability of R,N,C Stars\protect\\
 from HIPPARCOS and AAVSO Data}

\author{MICHEL GRENON}
\institute{Geneva Observatory, Sauverny, Switzerland}

\author{JANET A. MATTEI}
\institute{American Association of Variable Star Observers\\
Cambridge MA, U.S.A.} 

\author{LAURENT EYER}
\institute{Geneva Observatory, Sauverny, Switzerland}

\author{GRANT FOSTER}
\institute{American Association of Variable Star Observers\\
Cambridge, MA 02138, U.S.A.}

\end{opening}

\begin{document}

\begin{abstract}
Accurate photometry was obtained for all programme stars during the 3.3-year
HIPPARCOS mission. The final observing programme included several hundred 
Mirae (M), semiregular (SR) long-period and irregular (L) variables. A detailed
calibration of the aging of the optics allowed the evaluation of
very precise magnitudes over the whole range of star colours. 

\vspace*{0.2cm}
Since the time coverage of the satellite observations was not sufficient
to describe the behaviour of M, SR, or L type variables, smooth curves were
fitted statistically to the dense AAVSO observations. These
curves were then transformed to the HIPPARCOS system in order to complement
the HIPPARCOS photometry and thus produce precise light curves with fuller time
coverage, for a set of several hundred late-type variables, including most
Carbon stars brighter than V = 12.4 at minimum luminosity.

\vspace*{0.2cm}
A preliminary discussion of the behaviour of C stars, as observed from the
space in the broad Hp band, is given. 

\end{abstract}

\vspace*{-0.5cm}
\section{The LPVs in HIPPARCOS programme}

During the compilation of the HIPPARCOS Input Catalogue (HIC), special
attention was given to obtain a uniform all sky coverage of late type
variables, including the poorly known ones in the Southern hemisphere.
The inclusion in HIC of long period variables (LPVs) was possible only for
those brighter than HIPPARCOS magnitude Hp 12.5 (the detection threshold) 
at least 80 \% of their cycle. One of prerequisites of the HIPPARCOS mission
was that the brightness of the targets needed to be known in advance to
allocate the appropriate observing time. However LPVs are not strictly
periodic in their amplitudes, phases, and even periods. Thus the
prediction of the brightness and of the observability windows (time
intervals when Hp was $<$12.4) could not be achieved without performing
complementary ground-based observations before and during the mission on 
about 340 LPVs. The responsibility of monitoring the HIPPARCOS LPVs was
taken by the AAVSO, both by continuing long term observations and adding
new variables to the AAVSO observing programme (Mattei 1988). About one
million longterm AAVSO observations, together with about 70000 yearly
continuing observations were used to prepare and refine the ephemerides
produced by the variable star coordinator at Montpellier, France in
collaboration with the AAVSO.

\vspace*{0.2cm}
\section{The photometric reduction of LPVs}

The main-mission photometry was performed in the wide Hp band extending
from 380 to 900 nm. Due to irradiation by energetic solar and cosmic
particles, the transmission of the detection chain suffered a severe
wavelength-dependent deterioration during the mission. The standard Hp
system was re-defined for an epoch near mid-mission. A photometric reduction
to a subset of 22000 standard stars allowed to fix the zero point of the Hp
magnitude scale better than 0.001 mag twice per day. The very red
stars were a difficult case for the reduction to a standard Hp system. The
reddest non-variable standards have V-I colour less than 1.8 whereas the
mean V-I of most LPVs lies in a range from 2 to 6 magnitudes. Note that the
dominant flux for late type Carbon stars is emitted in the 700 to 900 nm
domain. The early reduction algorithms were polynomial relations between the
instrumental magnitudes and the standard Hp$_{s}$ as function of B-V. This
approach failed to model the aging effects for late type stars, especially
M and S type giants, inducing spurious long term drifts and short term
flickering when the reduction relations were extrapolated to red variables.

The technique to accurately define the chromatic aging from red variables,
i.e. $\delta$Hp versus V-I as a function of time, was accomplished by
forcing Hp - V$_{AAVSO}$ to be constant, at given light-curve phase and
V$_{AAVSO}$, throughout the mission. The $\delta$Hp /(V-I) relation is a non-
linear function of V-I. For reduction purposes, a linear pseudo-index was 
defined as given in Table 1.3.2 of The Hipparcos and Tycho Catalogues
(HIP), Vol. 1 (ESA 1997). This linearization procedure is nearly exact 
for stars with V-I less than 2, but may leave reduction residuals of the
order of few percent on individual magnitudes.

Simultaneous observations, visual by AAVSO observers and photoelectric, with
CCD camera and classical Geneva photometer, were performed to tie the Hp, the
$V_{CCD}$ and $V_{G}$, $V_{J}$ scales for red semiregular and Mira type LPVs. 
For M,S,C stars a unique relation exists between Hp and the Johnson $V_{J}$, 
as function of V-I. Note that M and C stars have a very distinct behaviour 
(Hp-V)/(B-V) or (Hp-$V_{T}$)/($B_{T}$-$V_{T})$ plane, see Fig. 1.3.7 in 
The Hipparcos and Tycho Catalogues (HIP), Vol. 1 (ESA 1997).

 The adopted relation Hp-V versus V-I from the Cousin's system, is given
below:\\

\begin{tabular}{llrrrrrrrrr}

V-I  & : &   2.00 & 2.50 & 3.00 & 3.50 & 4.00 & 4.50 &  5.00 &  5.50 &  6.00\\
\hline
Hp-V & : &   0.08 & 0.02 & -.09 & -.28 & -.53 & -.81 & -1.10 & -1.38 & -1.66
\end{tabular}

\vspace*{0.2cm}
\section{The light-curves of Carbon stars}

All R,N,C type stars observed by HIPPARCOS turned out to be variable either 
irregular, semi-regular or nearly periodic. Since $\lambda eff_{Hp}$ is 
larger  than $\lambda eff_{V}$, the amplitude in the Hp band is generally
smaller than that in B and V bands. This behaviour is illustrated in
Fig.1 for HIP 59844, the pulsating star BH Cru.

\begin{figure}
\label{VJHP_59844}
\setlength{\epsfxsize}{10cm}
\centerline{\epsffile{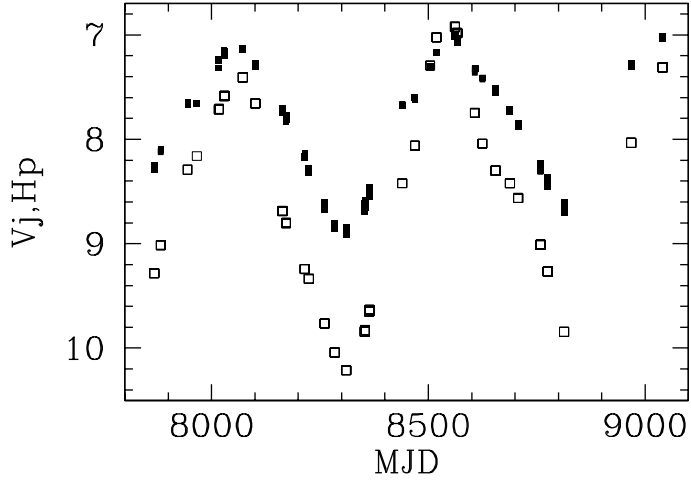}}
\caption{The light-curve of HIP 59844, of spectral type SC4,5-8e, in Hp
magnitude (filled squares) and in visual V magnitude (open squares) as 
deduced from AAVSO observations.}
\end{figure}

The ratio Q = $A_{Hp}/A_{V}$ is around 0.7 for early type C and M giants. It
shows a slight decrease for the reddest stars, see Fig.2.

\begin{figure}
\label{Q_aHV}
\setlength{\epsfxsize}{10cm}
\centerline{\epsffile{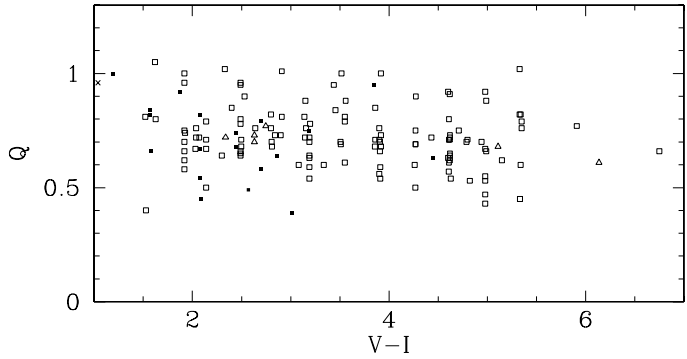}}
\caption{Q, the ratio between the Hp amplitude and the V amplitude, as a
function of the V-I colour for red variables. Symbol caption: open squares:
M type; open triangles: S type; filled squares: C type; crosses: K and M 
supergiants.}
\end{figure}

Large amplitude periodic variables often show nearly sine folded light-curves,
 e.g. HIP 109089 in Fig.3.

\begin{figure}
\label{F_Curves}
\setlength{\epsfxsize}{12cm}
\centerline{\epsffile{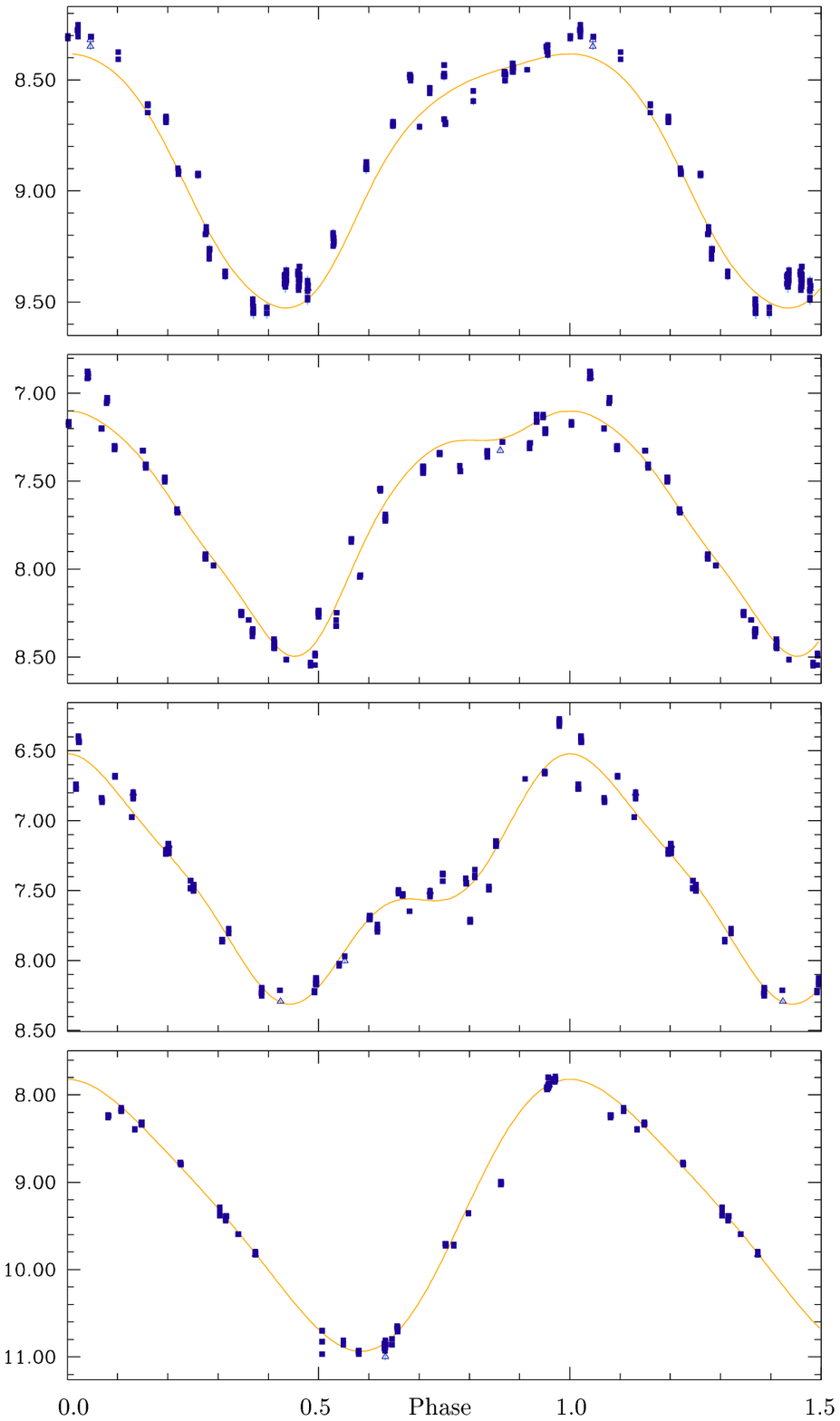}}
\caption{Folded light-curves of typical C-type periodic variables monitored
over 3 to 4 cycles. From top to bottom: HIP 26753, C0e, P=326d ; HIP 99653  
C5II, P=431d ; HIP 106583 C6II,  P=486d ; HIP 109089 C9e, P=436d. Error bars
are smaller than the symbol size The dotted line is the adopted best fit used
to derive the epoch and the magnitudes at brightness extrema.}
\end{figure}

The bump near phase 0.7 is present in many light-curves as shown for HIP
106583, 99653 or 26753. Semi-regulars and stars pulsating on the first
overtone show rather noisy folded light curves due to the their varying
amplitude and not so periodic behaviour. For small amplitude irregulars,
the time coverage during the HIPPARCOS mission was sufficient to describe
their behaviour in terms of peak to peak amplitudes and time scales for
variations, cf. Eyer et al.(1997).

The Hp amplitudes show two regimes for periodic R,N,C variables. R stars and 
part of C,N stars show a linear relation between the period and the amplitude
which may be expressed as $A_{Hp}$ = 1.3E-03$*$Period. For the classical C-rich
Mirae with typical $A_{Hp}$ in the range 1.2 to 2.4 mag, the amplitude 
shows little if any dependence on the period, ranging between 200 and 480
days. The global behaviour is displayed in Fig.4.

\begin{figure}
\label{RNC_amp}
\setlength{\epsfxsize}{10cm}
\centerline{\epsffile{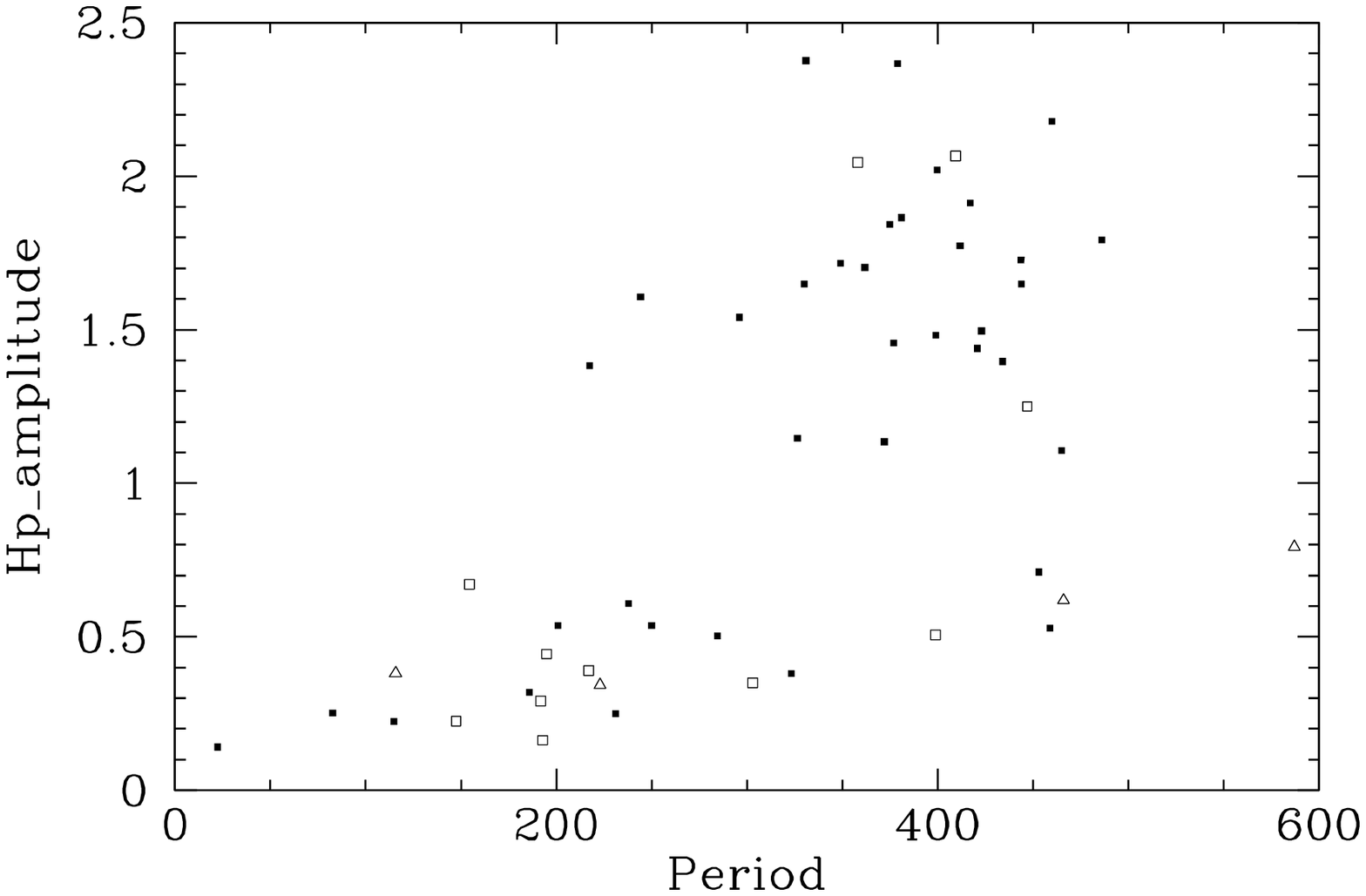}}
\caption{The relation between the amplitude in Hp band and the period, for 
periodic red variables of type C (filled squares), N(open squares) and R(open
triangles).}
\end{figure}

\vspace*{0.2cm} 
\section{The HIPPARCOS-AAVSO light-curves}

The visual estimates are obtained by interpolating the brightness of the 
variable star using a series of reference stars of known magnitude in its
field. The difference between the response of the eye and of the $V_{J}$ band
leads to an offset between the visual and the photoelectric $V_{J}$, 
proportional to the colour difference between the comparison stars and the
red variable. 

The monitoring by AAVSO observers generally produces a dense time coverage
but because observations of different observers are combined, and because
the accuracy of the individual observation is only between 0.1 and 0.3,
the light curves are noisy.  Thus, average light curves were obtained by
fitting curves to the individual observations by Fourier, polynomial or
quintic splines methods. These fitted curves were then transformed to Hp
magnitudes and HIPPARCOS photometry was then superimposed to them. 

The difference Hp-V is a function of the star's temperature and of the
circum- and inter- stellar extinction. Since the colour change as function
of the phase is generally unknown, the technique to reduce AAVSO
magnitudes to Hp is to plot $Hp-V_{AAVSO}$ versus $V_{AAVSO}$. For LPVs
the relation is often S-shaped as shown for the C0ev type star HIP 4284,
see Fig.5.

\begin{figure}
\label{VH_4284}
\setlength{\epsfxsize}{10cm}
\centerline{\epsffile{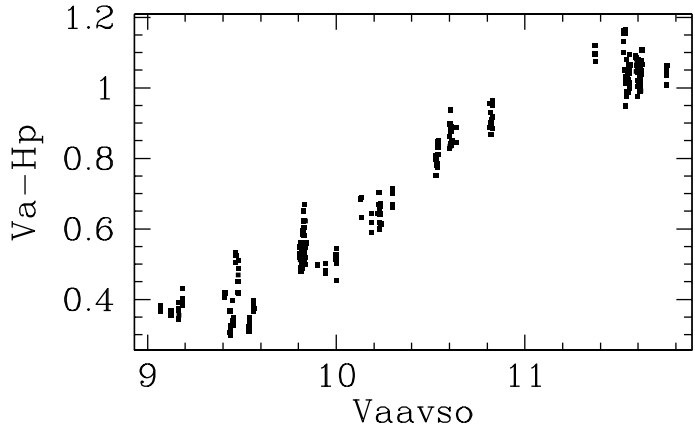}}
\caption{Example of non-linear relation $V_{AAVSO}$-Hp versus $V_{AAVSO}$ used
to transform visual light-curves into Hp light-curves.}
\end{figure}

The fine structure of $Hp-V_{AAVSO}$/$V_{AAVSO}$ diagrams depends mainly on 
the Teff and log g variations and on the corresponding absorption changes 
due to TiO, VO or CN, C2, SiC molecular bands. Emission lines and dust
extinction introduce departures from the mean relation. 

Although distinct relations seem to exist for rising and falling parts of the
light-curve, a unique third degree polynomial was used to transform fitted 
AAVSO magnitudes to Hp magnitudes with an uncertainty of 0.1- 0.2 mag in most
cases.This uncertainty is generally small compared to the peak-to-peak Hp
amplitude. Errors on comparison stars magnitudes are automatically corrected
by this process.

The Atlas of HIPPARCOS-AAVSO light-curves, part B (ESA 1997), contains 
274 stars in common. Here we show few representative cases of C stars 
light-curves which would be difficult to interpret with HIPPARCOS 
data alone. This is especially true for RCB variable RY Sgr, where the 
main minima were missed due to the peculiar HIPPARCOS time sampling, 
and the semiregular variable V Hya with a 530-day and over a 6000-day 
period where HIPPARCOS observations were obtained while the star was 
slowly fading to the minimum of its longer period, see Fig.6.

\begin{figure}
\label{L_Curves}
\setlength{\epsfxsize}{12cm}
\centerline{\epsffile{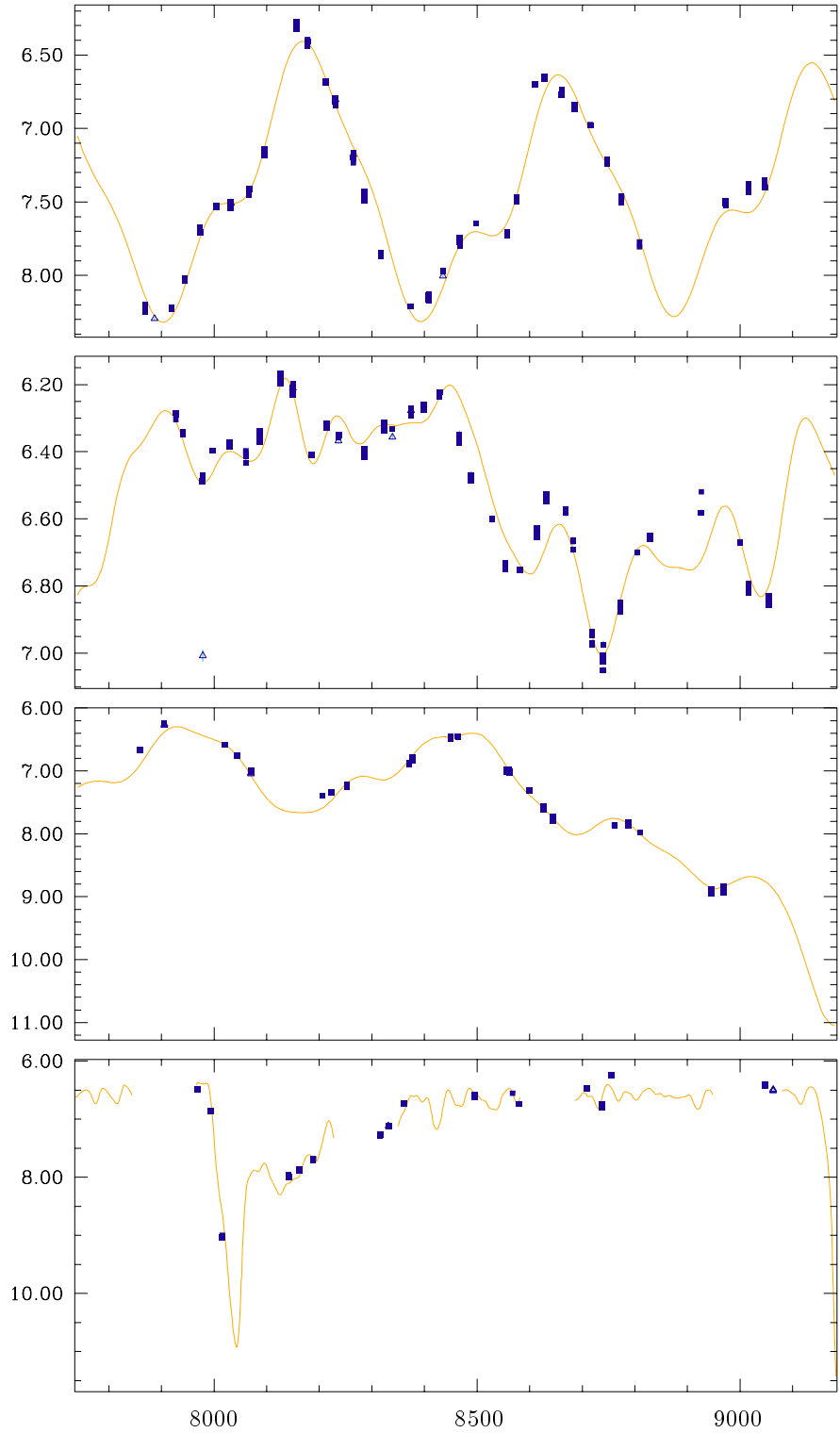}}
\caption{Light-curves of C-type Mira and
semiregular variables. From top to bottom: HIP 106583, C6II, S Cep; HIP
63152 C7I, RY Dra; HIP 53085 C9I, V Hya; HIP 94730 Cp, RY Sgr, an RCB
variable. Error bars are smaller than the symbol size.} \end{figure}

\section{Acknowledgements}

We sincerely thank variable star observers around the world whose
dedicated observations provided vital support in the observations of LPVs.
We gratefully acknowledge the support of NASA under grant NAGW-1493 
which made it possible for the AAVSO to provide data support to the 
HIPPARCOS mission and the Swiss National Science Foundation for its support 
to activities at Geneva Observatory.

\section{References}
European Space Agency (ESA). 1997, in {\it The Hipparcos and Tycho
 Catalogues}, ESA SP-1200, Vol. 1 and 12, The Netherlands.\\
Eyer, L. and Grenon M. 1997, in {\it Hipparcos - Venice '97, 13-16 May,\\
 Venice, Italy}, ESA SP-402 (July 1997), 467. \\
Mattei, J.A. 1988, in {\it Scientific Aspects of the Input Catalogue \\
Preparation II, Sitges, 25-29 January}, J. Torra, C. Turon Eds., 376.
\end{document}